\documentclass[%
 reprint,
 amsmath,amssymb,
 aps,
prb,
]{revtex4-2}

\usepackage{graphicx}
\usepackage{dcolumn}
\usepackage{bm}
\usepackage[utf8]{inputenc}
\usepackage{lipsum}
\usepackage{color}
\usepackage{wrapfig}

\begin{document}

\preprint{APS/123-QED}

\title{Competition of trivial and topological phases in patterned graphene based heterostructures}

\author{Zoltán Tajkov}
\affiliation{
 Department of Biological Physics, ELTE E{\" o}tv{\" o}s Loránd University, P{\'a}zmány P{\'e}ter s{\' e}t{\' a}ny 1/A, 1117 Budapest, Hungary
}
 \email{novidad21@caesar.elte.hu}

\author{János Koltai}
\affiliation{
 Department of Biological Physics, ELTE E{\" o}tv{\" o}s Loránd University, P{\'a}zmány P{\'e}ter s{\' e}t{\' a}ny 1/A, 1117 Budapest, Hungary
}

\author{József Cserti}
\affiliation{
 Department of Physics of Complex Systems, ELTE E{\" o}tv{\" o}s Loránd University, P{\'a}zmány P{\'e}ter s{\' e}t{\' a}ny 1/A, 1117 Budapest, Hungary}

\author{László Oroszlány}
\affiliation{
 Department of Physics of Complex Systems, ELTE E{\" o}tv{\" o}s Loránd University, P{\'a}zmány P{\'e}ter s{\' e}t{\' a}ny 1/A, 1117 Budapest, Hungary; MTA-BME Lendület Topology and Correlation Research Group, Budafoki {\'u}t 8., H-1111 Budapest, Hungary}%

\date{\today}

\begin{abstract}

We explore the effect of mechanical strain on the electronic spectrum of
patterned graphene based heterostructures. We focus on the competition of
Kekulé-O type distortion favoring a trivial phase and commensurate Kane-Mele
type spin-orbit coupling generating a topological phase. We derive a simple
low-energy Dirac Hamiltonian incorporating the two gap promoting mechanisms and
include terms corresponding to uniaxial strain. The derived effective model
explains previous \emph{ab initio} results through a simple physical picture.
We show that while the trivial gap is sensitive to mechanical distortions, the
topological gap stays resilient.

\end{abstract}

\maketitle

\section{\label{sec:level1}Introduction}

Spin-orbitronics is a promising new paradigm of nano-electronics that utilizes
the coupling of spin and orbital degrees of freedom of charge carriers
\cite{awschalom2009trend,kuschel2015spin}, enabling the non-magnetic control of
spin currents. The exceptionally long spin dephasing time makes graphene an ideal
template material for spintronics applications
\cite{novoselov_electric_2004,neto2009electronic,han2014graphene,huertas2009spin}.
Kekulé type patterned bond texture has potentially a dramatic impact on the
electronic spectrum of graphene  { \cite{gamayun2018valley,
hou2007electron, chamon2000solitons, Kekule_BiSe}}. Specifically Kekulé-O type
distortion scatters carriers between the two momentum space valleys of graphene
resulting in the appearance of a band gap \cite{chamon2000solitons}.
{ Mechanical distortions have also a considerable effect on the
electronic states of graphene. Recently several experimental groups reported
the controlled manipulation of strain fields in graphene based heterostroctures
\cite{strain_PrezGarza_2014,strain_Jiang_2017,strain_Guan_2017,
strain_Goldsche_2018,strain_Wang_2019,strain_Wang_2020}.} Andrade and
coworkers have recently studied the effect of mechanical strain on the spectrum
of Kekulé patterned samples \cite{andrade2019valley}. They showed that strain
substantially influences the gap generated due to a Kekulé-O pattern. As strain
increases the gap closes pushing the sample from an insulator to a semimetal.
Crucially these previous studies did not consider other gap generation mechanisms
which may arise in samples compatible with Kekulé-O texture.
In our previous works we showed that in graphene based heterostructures,
utilizing ternary bismuth tellurohalides, Kekulé distortion is accompanied by
strong induced spin-orbit interaction (SOI) \cite{tajkov2019uniaxial,
tajkov2019topological}. In these structures a competition between two gap
generation mechanisms arise. On the one hand Kekulé-O texture favors a trivial
band gap, while the induced spin-orbit interaction drives a topological Kane-Mele
type gap \cite{Kane_Mele}. Consequently it has been found that mechanical
distortions can drive the system from a trivial insulator to a topological phase.

\begin{figure}[t!]
\includegraphics[width=0.8\columnwidth]{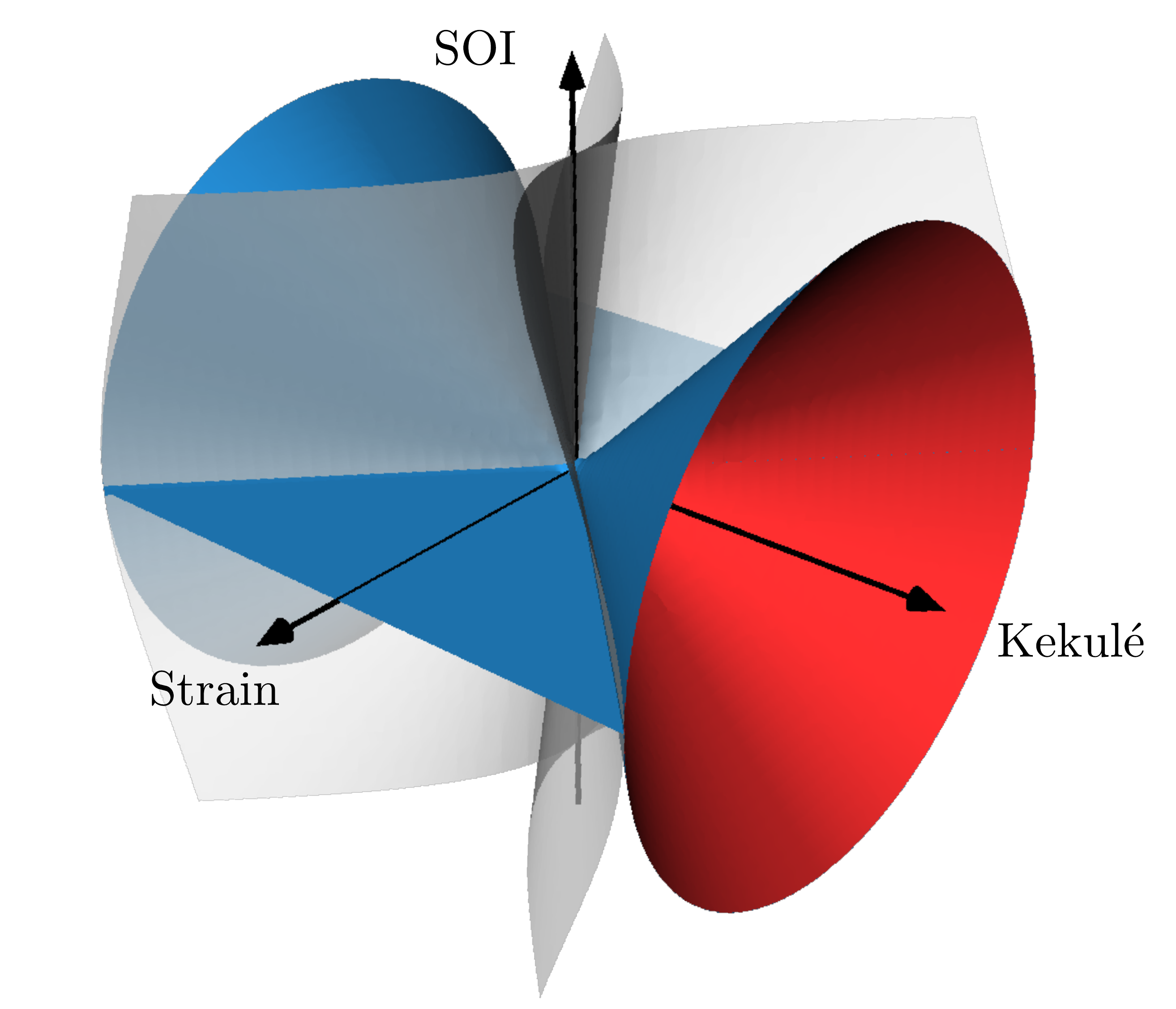}
\caption{{The phase diagram of the considered model based on Eq.
\eqref{eq:2} and the discussion below. Parameter set from the inside of the cones
depicts a trivial gap, while outside of the cones the system is topological.
The band gap is closed on the solid surfaces. Between the gray opaque shape and
the cones the gap is located at zero momentum,
everywhere else it is shifted away to a finite value.\\
(color online)}}
\label{fig:3D}
\end{figure}

In this manuscript we present a simple effective low-energy Dirac model that
captures the essence of this mechanically controlled topological quantum phase
transition.
Our findings are summarized in a three-dimensional phase diagram extracted from
the investigated model as depicted in Fig.~\ref{fig:3D}. Inside the cones the
Kekulé distortion outpowers the spin-orbit coupling, therefore the system is a
trivial insulator, while outside of the cones, spin-orbit coupling is dominant,
subsequently the gap is topological. Solid surfaces denote the phase
boundaries, where the gap is closed and the system is metallic.
We first introduce the low-energy Hamiltonian and then determine the low-energy
spectrum. Based on analytical properties of the spectrum we elucidate how
mechanical distortions influence the competition of the two types of gap promoting
effects.  {In Appendix A we give a detailed derivation of the
low-energy Hamiltonian based on a tight-binding model inspired by previous first
principles calculations. In Appendix B we outline the procedure we used to
classify the topological phases of the investigated system.}

\section{Model and Results}%
\label{model}

\begin{figure}[h!]
\includegraphics[width=1\columnwidth]{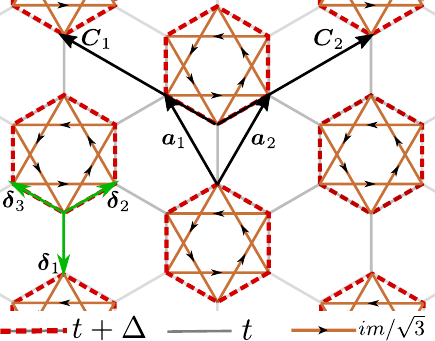}
\caption{Schematic structure of the unstrained model. The pale gray hexagons
represent the unadulterated  graphene substrate. Each pale gray bond corresponds
to a hopping amplitude $t$.
The red dashed lines denote the Kekulé-O type bond alteration with a characteristic
strength of $\Delta$. Each gray bond with a red dashed line on top corresponds to
a hopping amplitude $t+\Delta$.
Brown lines connecting next-nearest neighbor sites symbolize the induced Kane-Mele type
spin-orbit coupling. Each brown line corresponds to a spin dependent hopping amplitude
${i}m$, in the direction of the arrows. For more details see the Appendix.\\
(color online)
}
\label{fig:1}
\end{figure}

Let us consider the tight-binding model of a patterned graphene lattice as
depicted in Fig.\:\ref{fig:1}. Each first nearest neighbor bond carries a hopping
amplitude of $t$. Kekulé-O distortion is taken into account through an additional
$\Delta$ hopping term supplementing each bond around every third hexagon. These
terms will be responsible for opening a trivial gap \cite{chamon2000solitons}.
On the same hexagons the Kekulé-O distortion is active we also introduce a
next-nearest neighbour spin dependent hopping. For spin up electrons this
corresponds to a hopping amplitude ${i}m$ in the direction of the bond vectors
denoted by arrows, while $-{i}m$ for spin down particles. This term, adopted from
the Kane-Mele model, promotes a topological gap \cite{Kane_Mele}.
{  Also note that the considered spin-orbit coupling preserves the $z$
component of the spin operator, thus the up and down spin orientations can
effectively be treated separately.} Besides the considered bond texture discussed
above we also incorporate a uniaxial in-plane strain in our description. We take
into account the linear distortion of the lattice vectors and the exponential
renormalization of all hopping terms.

{ 

Since the considered model is diagonal in the spin degree of freedom its
$\mathbb{Z}_2$ topological index, characterizing time reversal symmetric systems,
is given by the parity of the total Chern number calculated for one spin component
of the occupied bands \cite{Fukui2005,fukui2007quantum,asboth_short_2016}. In
Appendix B we briefly summarize the procedure to obtain the Chern number of the
investigated system.
}

The following simplified low-energy Hamiltonian captures the most important aspects
of the electronic structure in the vicinity of the distorted Dirac cones of
graphene (for derivation of the model see  {in Appendix A}):

\begin{align}
\begin{split}
H_\mathrm{eff} &=-v\left[\boldsymbol{\sigma}\boldsymbol{A}\otimes\tau_{z}
\otimes s_0+\boldsymbol{\sigma}\boldsymbol{k}\otimes\tau_{0}\otimes s_0\right] \\
&+\Delta\sigma_{z}\otimes\tau_{x}\otimes s_0-m \sigma_{z}\otimes\tau_{0}\otimes s_z,
\end{split}
\label{eq:1}
\end{align}

where $\boldsymbol{k}$ is the momentum, $v=3a_\mathrm{cc}/2t$ is the Fermi velocity
(in units of $\hbar$) with $a_\mathrm{cc}=1.42\, \mathrm{\AA}$ the equilibrium
carbon-carbon bond distance \cite{cooper_experimental_2012}. The $\boldsymbol{\sigma}=
(\sigma_x, \sigma_y)$, $\tau_i$ and $s_i$ are the Pauli matrices. $\sigma_i$ are
acting on the sublattice,  $\tau_i$ on the valley degree of freedom while $s_i$
act on the real spin.
The pseudovector potential $\boldsymbol{A}$ describes mechanical distortions,
specifically for the case of uniaxial in-plane strain its components are
\cite{strainAndoCNT,strainManesGR,guinea2010energy,masir2013pseudo}:
\begin{equation}
\boldsymbol{A}	=\frac{\beta}{2a_\mathrm{cc}}\varepsilon\left(1+\rho\right)
\left(\begin{array}{r}
\cos2\theta\\
-\sin2\theta
\end{array}\right),
\end{equation}
where $\beta\approx 3$ is the Grüneisen parameter \cite{naumis2017electronic,
reich2000shear,gilvarry1956gruneisen} that modulates the hopping terms of the
tight-binding model as strain changes the intercarbon distance due to lattice
deformations, $\varepsilon$ is the magnitude of the distortion, $\rho$ is the
Poisson ratio, estimated to be $\rho \approx 0.165$ for graphene
\cite{botello2018toward,faccio2009mechanical,scarpa2009effective} while $\theta$
is the angular direction of the strain with respect to the $x$ axis.

Since Hamiltonian (\ref{eq:1}) is diagonal in the proper spin degree of freedom
each spin species can be treated separately. Diagonalizing (\ref{eq:1}) yields
the same spectrum for both spin orientations:

\begin{widetext}
\begin{equation}
E\left(\boldsymbol{k} \right)=\pm \sqrt{\xi^{2} +\Delta^{2}  +   v^{2}|
\boldsymbol{k}|^2 + m^{2} \pm 2 \sqrt{\xi^{2}   \Delta^{2} + \left(v^2
\boldsymbol{A}\cdot \boldsymbol{k} \right)^2 + \Delta^{2} m^{2}}},
\label{eq:2}
\end{equation}
\end{widetext}
where we introduce
\begin{equation}
\xi = v\frac{\beta}{2a_\mathrm{cc}}\varepsilon\left(1+\rho\right).
\end{equation}

\begin{figure}[t!]
\includegraphics[width=1\columnwidth]{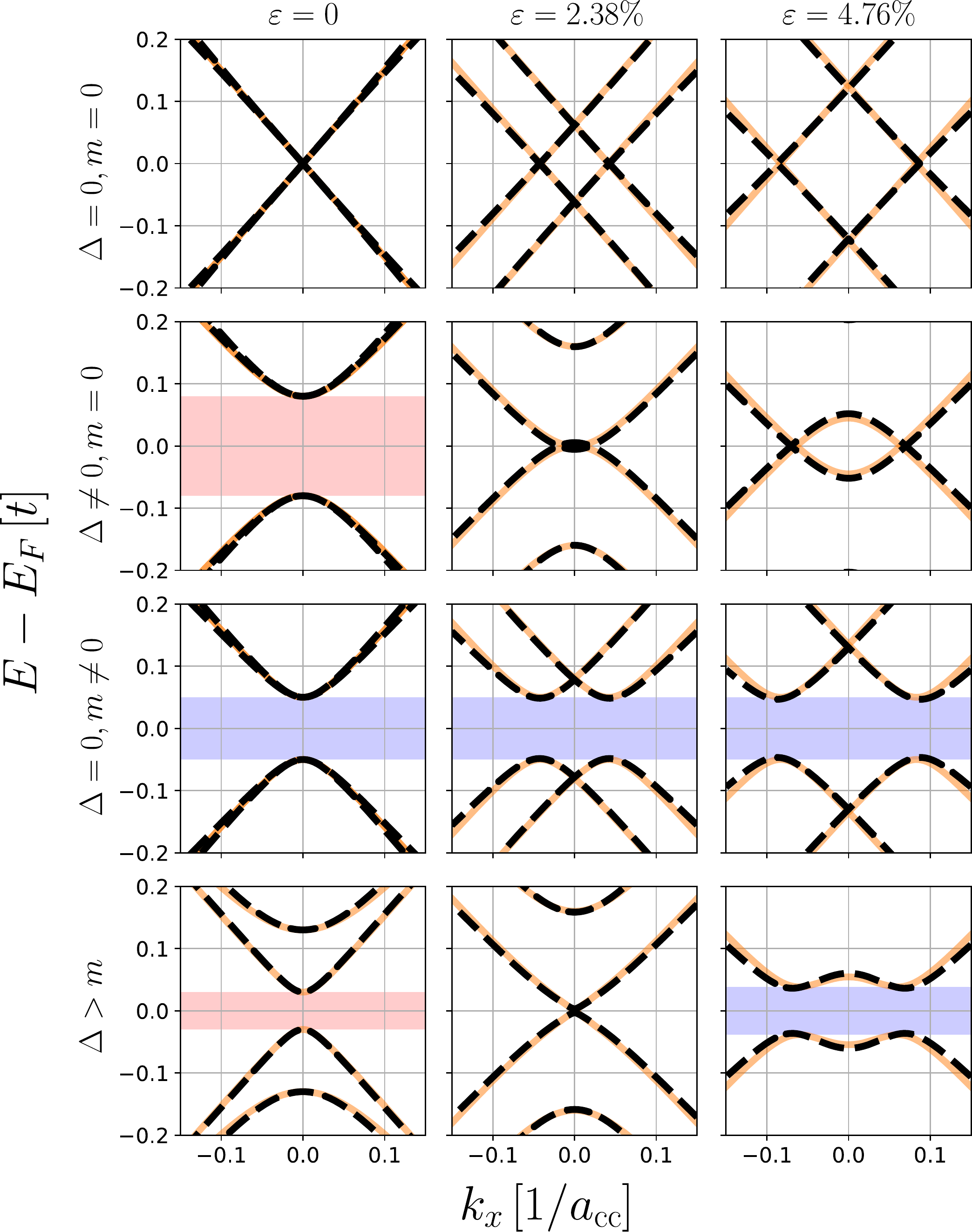}
\caption{The low-energy spectrum of the considered tight-binding model
\eqref{eq:fullham}, denoted by solid orange lines, and the effective Dirac
Hamiltonian \eqref{eq:1}, depicted with dashed black lines, for various parameters.
Topological gaps are indicated by blue opaque coloring, while trivial gaps are
highlighted with red. In all panels $m=0.04t$ and $\Delta=0.08t$ unless indicated
otherwise, for all cases $\theta=0$.\\
(color online)}
\label{fig:4abra}
\end{figure}

The low-energy spectrum of the considered tight-binding model \eqref{eq:fullham}
and the effective Dirac Hamiltonian \eqref{eq:1} for various model parameters is
shown in Fig.\,\ref{fig:4abra}. In the first row we show a clean graphene sample
under strain. As the magnitude of the mechanical distortion increases the two
Dirac cones shift apart.
In the second row a trivial gap opened by Kekulé pattern is closed by an ever
increasing strain driving the system into a semimetallic phase.
On the contrary, as it can be observed in the third row the topological gap opened
by the Kane-Mele term is insensitive to the strength of distortion.
In the last row both gap generation mechanism are active and $\Delta>m$
resulting in a trivial gap for the unstrained case. As strain is increased the
gap is closed and reopened but now with a topological flavor.

\begin{figure}[t!]
\includegraphics[width=1\columnwidth]{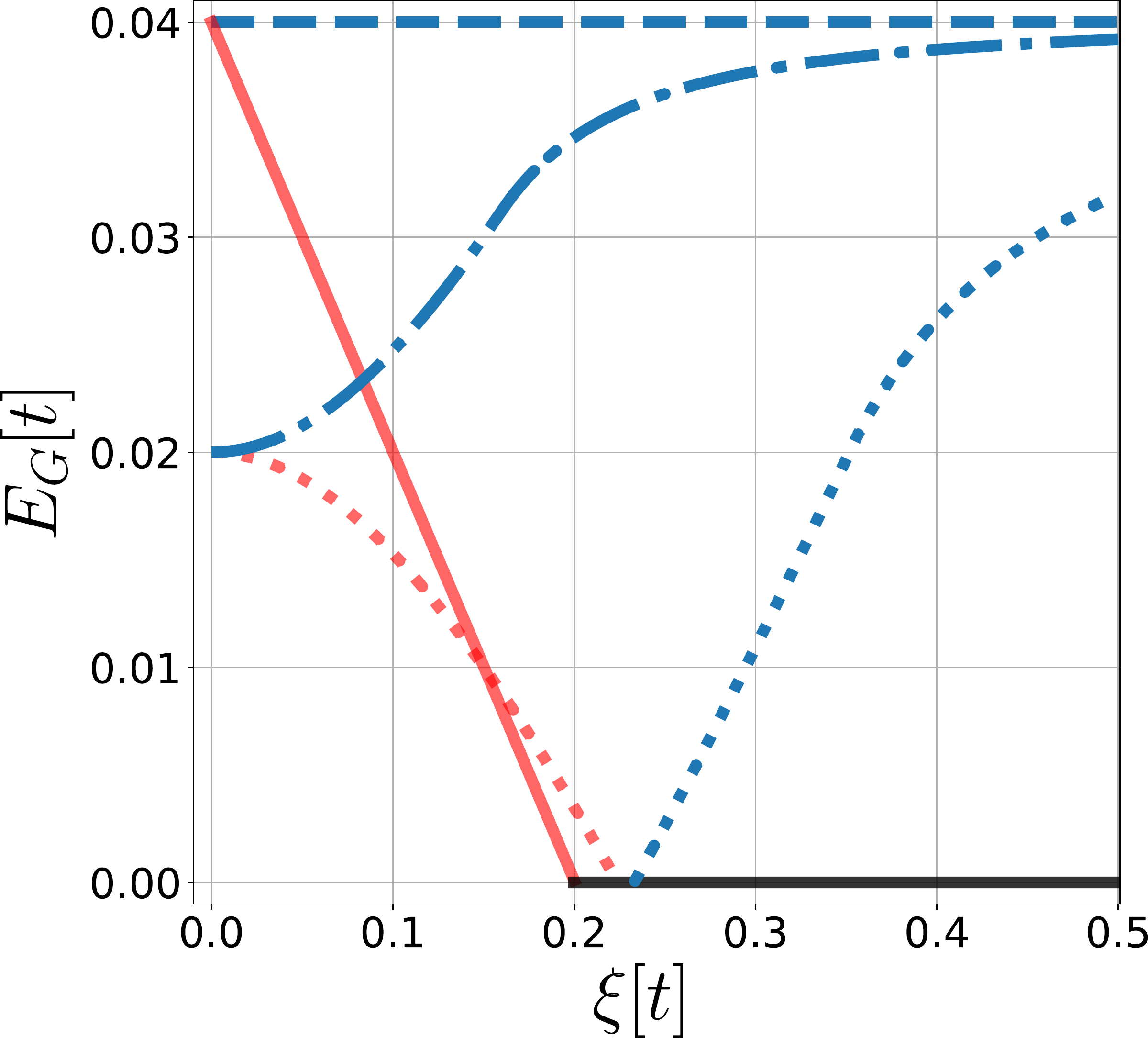}
\caption{
The magnitude of the gap obtained from the effective
Dirac Hamiltonian is shown as the function of uniaxial strain for various model
parameters. Red color indicates that the gap is trivial, while blue coloring
shows that the gap is topological.
The dashed line corresponds to $\Delta=0$ and $m=0.04t$. The solid line shows the
case when $\Delta=0.04t$ and $m=0$. The dashdotted line was obtained using
$\Delta=0.01t$ and $m=0.02t$ while in the case of the dotted curve it was
$\Delta=0.03t$ and $m=0.02t$. \\
\small{(color online)}}
\label{fig:competition}
\end{figure}

Analyzing the spectrum, depicted in Fig.\:\ref{fig:4abra}, the conditions for
sustaining a gap can be discerned.
We find, that if the applied mechanical distortion is constrained as
\begin{equation}
\label{eq:strain}
2\xi^2 \leq \sqrt{\Delta^2\left( \Delta^2+4m^2 \right)}+\Delta^2,
\end{equation}
then the band extrema are at $\boldsymbol{k}=(0,0)$ and the magnitude of
the gap is
\begin{equation}
E_\mathrm{G}=2\sqrt{\xi^{2} +\Delta^{2}  + m^{2} - 2 \sqrt{\xi^{2}  \Delta^{2} +
\Delta^{2} m^{2} }}.
\label{eq:gap1}
\end{equation}

The gap closes if $\xi=\sqrt{\Delta^2-m^2}$ at which point valence and conduction
band touch at a single point.
If condition \eqref{eq:strain} is unsatisfied the band extrema are shifted to a
finite momentum, the magnitude of the gap in this case is:

\begin{equation}
E_\mathrm{G}=2m\sqrt{1-\frac{\Delta^2}{\xi^2}}
\label{eq:gap2}
\end{equation}
We note that the observations made above are insensitive to the direction of
strain \emph{i.e.}\:$\theta$, similarly to previous results
\cite{naumis2017electronic,andrade2019valley}.

The impact of strain on the competition between the two topologically distinct
phases can be observed in Fig.~\ref{fig:competition}.
Here we plot the evolution of the magnitude and character of the gap as the
function of the strength of the applied mechanical distortion based on
\eqref{eq:gap1} and \eqref{eq:gap2}. If only the Kekulé term is active, shown
with solid lines, then the gap decreases linearly with increasing strain and
once it is closed the system remains gapless.
On the other hand if we only keep the next-nearest spin-orbit coupling, indicated
by dashed lines, the gap remains constant in the face of ever increasing strain.
If $\Delta>m$, shown with dotted lines, the original trivial gap closes and
a nontrivial gap opens as strain is increased. For $m>\Delta$, denoted by a
dash-dotted line, an initial smaller topological gap is increased by the application
of mechanical distortions.

\section{Conclusion}

We studied the effects of uniaxial strain upon the electronic properties of a
patterned graphene lattice hosting a Kekulé-O textured bond alternation and
a commensurate Kane-Mele type spin-orbit coupling.
Based on a tight-binding model we distilled an effective, low-energy
Dirac Hamiltonian. Analyzing the spectrum of the model we explored the various
gaped phases present in the sample.
 We found that while the trivial gap favored
by the Kekulé-O distortion is destroyed by the strain, the topological gap
generated by the considered spin-orbit term remains resilient.
This observation can be understood from the following reasoning.
The applied mechanical distortion moves the two Dirac cones of graphene from
their initial position. The Kekulé term scatters particles between valleys, and
hence it can only open a gap if the Dirac cones are aligned. The Kane-Mele term
on the other hand does not mix valleys and the gap opened through it is
insensitive to the position of the Dirac cones.
These findings explain our previous \emph{ab initio} results \cite{tajkov2019uniaxial}.

\section*{Conflicts of interest}
There are no conflicts to declare.

\section*{Acknowledgements}

This work was financially supported by the  the Hungarian National Research,
Development and Innovation Office (NKFIH) via the National Quantum Technologies
Program 2017-1.2.1-NKP-2017-00001; grants no. K112918, K115608, FK124723 and K115575.
This work was completed in the ELTE Excellence Program (1783-3/2018/FEKUTSRAT)
supported by the Hungarian Ministry of Human Capacities. Supported by the ÚNKP-19-4
New National Excellence Program of the Ministry for Innovation and Technology. J.K.
acknowledges the Bolyai program of the Hungarian Academy of Sciences. We acknowledge
[NIIF] for awarding us access to resource based in Hungary at Debrecen.

\appendix
\section{}

In this Section, we present the derivation of our simplified low-energy Hamiltonian
(Eq.~\eqref{eq:1}). The system of interest is the Kekulé-O distorted graphene
lattice hosting next-nearest neighbor Kane-Mele type spin-orbit interaction in
the perturbed hexagons (See Fig.~\ref{fig:1}).
To study the impact of mechanical strain on the electronic spectrum, we introduced
uniform, planar strain,\textit{id est}, the strain is position independent. Thus
the displacement field of the atoms due to the deformation is given by:
$\boldsymbol{u}(\tilde{\boldsymbol{r}})=\boldsymbol{\varepsilon}\cdot
\tilde{\boldsymbol{r}}$, where $\tilde{\boldsymbol{r}}$ is the original position
of the cores. After deformation the position of the atoms are $\boldsymbol{r}=
\tilde{\boldsymbol{r}} +\boldsymbol{u}(\tilde{\boldsymbol{r}})=\left( \boldsymbol{I}+
\boldsymbol{\varepsilon}\right)\tilde{\boldsymbol{r}}$, where $\boldsymbol{I}$ is
the 2$\times$2 identity matrix and the deformation tensor $\boldsymbol{\varepsilon}$
can be parametrized as:

\begin{equation}
\boldsymbol{\varepsilon}=\varepsilon\left(\begin{array}{cc}
\mathrm{cos}^{2}\theta-\rho\mathrm{sin^{2}\theta} & \left(1+\rho\right)\mathrm{cos}\theta\mathrm{sin}\theta\\
\left(1+\rho\right)\mathrm{cos}\theta\mathrm{sin}\theta & \mathrm{sin^{2}\theta}-\rho\mathrm{cos}^{2}\theta
\end{array}\right),
\end{equation}
where $\varepsilon$ is the magnitude of the applied strain, $\theta$ is its angular
direction, with respect to the $x$ axis and $\rho$ is the Poisson ratio, which
relates the transverse strain to the axial component (estimated to be $\rho= 0.165$
for graphene).

We applied the above outlined formalism to include strain in our model, which can
be formulated as:

\begin{equation}
\label{eq:whole}
\hat{H}=\left(\begin{array}{cc}
\hat{h}_\uparrow & 0\\
0 & \hat{h}_\downarrow
\end{array}\right).
\end{equation}
Since Hamiltonian (\ref{eq:whole}) is diagonal in the proper spin degree of
freedom \cite{Kane_Mele} each spin species can be treated separately:

\begin{equation}
\label{eq:whole2}
\hat{h}_{\uparrow/\downarrow}=\hat{H}_\mathrm{Gr}+\hat{H}_\mathrm{Kek}\pm
\hat{H}_\mathrm{SOI},
\end{equation}
where $\hat{H}_\mathrm{Gr}$ describes the pristine graphene, $\hat{H}_\mathrm{Kek}$
is the Hamiltonian of the Kekulé distortion and $\hat{H}_\mathrm{SOI}$ incorporates
the patterned next-nearest-neighbor spin-orbit interaction.

First let us consider the clean, pristine graphene. The structure of graphene is
defined by a unit cell consisting of two equivalent atoms A and B with one $\pi$
orbital per carbon atom considered.  The lattice is spanned by the lattice
vectors: $\tilde{\boldsymbol{a}}_{1,2}	=a_{\mathrm{cc}}\frac{1}{2}\left(\mp\sqrt{3},\, 3\right)$,
where ${a_{\mathrm{cc}}}\approx 1.42 \, \mathrm{\AA}$ is the
unperturbed carbon-carbon distance in the lattice. We define single particle
states situated on sublattice $\alpha=A\,\mathrm{or}\,B$ as:

\begin{equation}
\left|\alpha,\underline{m}\cdot\underline{\boldsymbol{a}}\right\rangle =\left|
\alpha\right\rangle \left|\underline{m}\cdot\underline{\boldsymbol{a}}\right\rangle
\end{equation}
where $\underline{m}\cdot\underline{\boldsymbol{a}}$ runs over the atomic
positions as $\underline{m}\cdot\underline{\boldsymbol{a}}=m_{1}\boldsymbol{a}_{1}+m_{2}\boldsymbol{a}_{2}$
with $m_1$ and $m_2$ being integers. With this notation
the tight-binding Hamiltonian for strained graphene in real space is given by:
\begin{equation}
\label{eq:Hgr}
\begin{split}
\hat{H}_\mathrm{Gr}=t\sum_{\underline{m}} & d_1\left|B,\underline{m}\cdot
\underline{\boldsymbol{a}}\right\rangle \left\langle A,\underline{m}\cdot
\underline{\boldsymbol{a}}\right| \\
 + & d_2\left|B,\underline{m}\cdot\underline{\boldsymbol{a}}\right\rangle
 \left\langle A,\underline{m}\cdot\underline{\boldsymbol{a}}+\boldsymbol{a}_1
 \right| \\
 + & d_3\left|B,\underline{m}\cdot\underline{\boldsymbol{a}}\right\rangle
 \left\langle A,\underline{m}\cdot\underline{\boldsymbol{a}}+\boldsymbol{a}_2
 \right|+\text{h.c.},
\end{split}
\end{equation}
where $t\approx 2.7 \, \mathrm{eV}$ is the hopping parameter of unstrained
graphene and the $d_i$ factors describe the strain.
Hence the overlap integrals depend on the actual position of the atoms, the effect
of strain on the hopping terms can be taken into account with the following
factors \cite{turchi1998tight,pereira2009tight,neto2009electronic}:

\begin{equation}
\label{eq:strain_hopping}
d_{i}=\mathrm{e}^{-\beta\left(\frac{|\boldsymbol{\delta}_{i}|}{a_{\mathrm{cc}}}
-1\right)}, \qquad i=1,2,3,
\end{equation}
where $\beta$ is the Grüneisen parameter (estimated to be $\beta \approx 3$ for
graphene) and $\boldsymbol{\delta}_{i}$ are the bonding vectors pointing to one
of the three nearest-neighbor sites at a given $\boldsymbol{r}$, as shown in
Fig.\:\ref{fig:1}, $\tilde{\boldsymbol{\delta}}_{1}=a_\mathrm{cc}\left(0,-1
\right),\tilde{\boldsymbol{\delta}}_{2}=a_\mathrm{cc}/2\left(\sqrt{3},1 \right)$,
$\tilde{\boldsymbol{\delta}}_{3}=a_\mathrm{cc}/2\left(-\sqrt{3},1 \right)$.

In the next step, let us introduce the Kekulé-O distortion. The texture of the
ordering is visualized in Fig. \ref{fig:1}. The periodicity of the pattern differs
from the pristine graphene and can be written as: $\boldsymbol{C}_{1}=2
\boldsymbol{a}_{1}-\boldsymbol{a}_{2}$ and $\boldsymbol{C}_{2}=-\boldsymbol{a}_{1}
+2\boldsymbol{a}_{2}.$ We can formulate the Hamiltonian of the Kekulé-O texture
in real space as:

\begin{equation}
\begin{split}
\label{eq:kek}
\hat{H}_\mathrm{KeK}&={\Delta}\sum_{\underline{M}} d_{2} \left|B,\underline{M}\cdot \underline{\boldsymbol{C}}\right\rangle \left\langle A,\underline{M}\cdot \underline{\boldsymbol{C}}+\boldsymbol{a}_{2}\right| \\
& +   d_{1}\left|B,\underline{M}\cdot \underline{\boldsymbol{C}}+\boldsymbol{a}_{2}\right\rangle \left\langle A,\underline{M}\cdot \underline{\boldsymbol{C}}+\boldsymbol{a}_{2}\right| \\
& +   d_{3}\left|B,\underline{M}\cdot \underline{\boldsymbol{C}}+\boldsymbol{a}_{2}\right\rangle \left\langle A,\underline{M}\cdot \underline{\boldsymbol{C}}+\boldsymbol{a}_{2}+\boldsymbol{a}_{1}\right|\\
& +   d_{2}\left|B,\underline{M}\cdot \underline{\boldsymbol{C}}+\boldsymbol{a}_{1}\right\rangle \left\langle A,\underline{M}\cdot \underline{\boldsymbol{C}}+\boldsymbol{a}_{2}+\boldsymbol{a}_{1}\right|\\
& +   d_{1}\left|B,\underline{M}\cdot \underline{\boldsymbol{C}}+\boldsymbol{a}_{1}\right\rangle \left\langle A,\underline{M}\cdot \underline{\boldsymbol{C}}+\boldsymbol{a}_{1}\right| \\
& +   d_{3}\left|B,\underline{M}\cdot \underline{\boldsymbol{C}}\right\rangle \left\langle A,\underline{M}\cdot \underline{\boldsymbol{C}}+\boldsymbol{a}_{1}\right| + \text{h.c.},
\end{split}
\end{equation}
where $\underline{M}\cdot\underline{\boldsymbol{C}}$ runs over the atomic positions
as $\underline{M}\cdot\underline{\boldsymbol{C}}=M_{1}\boldsymbol{C}_{1}+M_{2}
\boldsymbol{C}_{2}$ with $M_1$ and $M_2$ being integers and $\Delta$ is the strength
of the Kekulé distortion. (\textit{We note that our definiton of $\Delta$ is
different from \cite{andrade2019valley}}).

The last part of our model is the next-nearest neighbor spin-orbit interaction.
In order to preserve time-reversal symmetry we write the SOI term as $im/\sqrt{3}$,
where $m$ is the strength of the interaction and has a real value.  We introduced
an extra $1/\sqrt{3}$ factor in the magnitude of the interaction, which will be
convenient later. The real space periodicity of this term is the same as in the
case of the Kekulé texture:

\begin{equation}
\begin{split}
	\hat{H}_\mathrm{SOI} &= \frac{im}{\sqrt{3}}\sum_{\underline{M}} d_{5}\left|B,\underline{M}\cdot \underline{\boldsymbol{C}}+\boldsymbol{a}_{2}\right\rangle  \left\langle B,\underline{M}\cdot \underline{\boldsymbol{C}}\right| \\
&+ d_{4}\left|B,\underline{M}\cdot \underline{\boldsymbol{C}}\right\rangle  \left\langle B,\underline{M}\cdot \underline{\boldsymbol{C}}+\boldsymbol{a}_{1}\right| \\
&+ d_{6}\left|B,\underline{M} \cdot\underline{\boldsymbol{C}}+\boldsymbol{a}_{1}\right\rangle  \left\langle B,\underline{M}\cdot \underline{\boldsymbol{C}}+\boldsymbol{a}_{2}\right| \\
&+ d_{5}\left|A,\underline{M}\cdot \underline{\boldsymbol{C}}+\boldsymbol{a}_{1}\right\rangle  \left\langle A,\underline{M}\cdot \underline{\boldsymbol{C}}+\boldsymbol{a}_{1}+\boldsymbol{a}_{2}\right| \\
&+ d_{4}\left|A,\underline{M}\cdot \underline{\boldsymbol{C}}+\boldsymbol{a}_{1}+\boldsymbol{a}_{2}\right\rangle  \left\langle A,\underline{M}\cdot \underline{\boldsymbol{C}}+\boldsymbol{a}_{2}\right| \\
&+ d_{6}\left|A,\underline{M}\cdot \underline{\boldsymbol{C}}+\boldsymbol{a}_{2}\right\rangle  \left\langle A,\underline{M}\cdot \underline{\boldsymbol{C}}+\boldsymbol{a}_{1}\right|+\text{h.c.}.
\end{split}
\end{equation}
The factors ($d_{4,5,6}$) of the hopping terms have a different form corresponding
to the next-nearest neigbor bonding vectors:
\begin{gather}
\label{eq:strain_hopping2}
d_{4}=\mathrm{e}^{-\beta\left(\frac{|\boldsymbol{a}_{1}|}{\sqrt{3}a_\mathrm{cc}}-1\right)},\\
d_{5}=\mathrm{e}^{-\beta\left(\frac{|\boldsymbol{a}_{2}|}{\sqrt{3}a_\mathrm{cc}}-1\right)},\nonumber\\
d_{6}=\mathrm{e}^{-\beta\left(\frac{|\boldsymbol{a}_{1}-\boldsymbol{a}_{2}|}{\sqrt{3}a_\mathrm{cc}}-1\right)}.\nonumber
\end{gather}
In order to compute the dispersion relation we must take the Fourier transform of
our Hamiltonian \eqref{eq:whole}. We define the transformation via the following
relations \cite{bena2009remarks}:
\begin{gather}
\label{eq:FT}
\left|B,\underline{m}\cdot\underline{\boldsymbol{a}}\right\rangle =\frac{1}{\sqrt{N}}\sum_{\boldsymbol{k}}\mathrm{e}^{i\boldsymbol{k}\underline{m}\cdot\underline{\boldsymbol{a}}}\left|B,\boldsymbol{k}\right\rangle, \\
\left|A,\underline{m}\cdot\underline{\boldsymbol{a}}\right\rangle =\frac{1}{\sqrt{N}}\sum_{\boldsymbol{k}}\mathrm{e}^{i\boldsymbol{k}\left(\underline{m}\cdot\underline{\boldsymbol{a}}+\boldsymbol{\delta}_{1}\right)}\left|A,\boldsymbol{k}\right\rangle \nonumber ,
\end{gather}
where $N$ is the number of the unit cells.

While calculating the Hamiltonian in $\boldsymbol{k}$-space is in principle
straightforward, we impart a couple of remarks. Using the definition in Eq.~\eqref{eq:FT}
we perform the Fourier transformation of the first term in Eq.~\eqref{eq:Hgr}.

\begin{gather}
\label{eq:FT2}
\sum_{\underline{m}}  d_{1}\left|B,\underline{m}\cdot\underline{\boldsymbol{a}}\right\rangle \left\langle A,\underline{m}\cdot\underline{\boldsymbol{a}}\right|\\
=\frac{d_{1}}{N}\sum_{\underline{m}}\sum_{\boldsymbol{k}',\boldsymbol{k}}\mathrm{e}^{i\boldsymbol{k}\underline{m}\cdot\underline{\boldsymbol{a}}}\mathrm{e}^{-i\boldsymbol{k}'\left(\underline{m}\cdot\underline{\boldsymbol{a}}+\boldsymbol{\delta}_{1}\right)}\left|B,\boldsymbol{k}\right\rangle \left\langle A,\boldsymbol{k}'\right|  \nonumber\\
=\frac{d_{1}}{N}\sum_{\underline{m}}\sum_{\boldsymbol{k}',\boldsymbol{k}}\mathrm{e}^{i\underline{m}\cdot\underline{\boldsymbol{a}}\left(\boldsymbol{k}-\boldsymbol{k}'\right)}\mathrm{e}^{-i\boldsymbol{k}'\boldsymbol{\delta}_{1}}\left|B,\boldsymbol{k}\right\rangle \left\langle A,\boldsymbol{k}'\right|.\nonumber
\end{gather}
Examine more closely the following term:
\begin{equation}
\label{eq:delta}
\sum_{\underline{m}}\mathrm{e}^{i\underline{m}\cdot\underline{\boldsymbol{a}}\left(\boldsymbol{k}-\boldsymbol{k}'\right)}
=\sum_{m_1=1}^{n_1}\sum_{m_2=1}^{n_2}\mathrm{e}^{i\left( m_1\boldsymbol{a}_1+m_2\boldsymbol{a}_2 \right)\Delta\boldsymbol{k}},
\end{equation}
where $n_1\cdot n_2=N$ and $\Delta \boldsymbol{k}=\boldsymbol{k}-\boldsymbol{k}'$.
We rewrite the $\Delta\boldsymbol{k}$ term as a linear combination of the reciprocal
lattice vectors: $\Delta\boldsymbol{k}=\Delta k_1 \cdot \boldsymbol{b}_1+\Delta
k_2 \cdot \boldsymbol{b}_2$, where $\Delta k_i=\frac{l}{n_i}$ with $l=0...n_i-1$.
Plugging this back to Eq. \eqref{eq:delta}:

\begin{gather}
\sum_{\underline{m}}\mathrm{e}^{i\underline{m}\cdot\underline{\boldsymbol{a}}\left(\boldsymbol{k}-\boldsymbol{k}'\right)}
=\sum_{m_1=1}^{n_1}\mathrm{e}^{2\pi i  m_1\Delta k_1}\sum_{m_2=1}^{n_2}\mathrm{e}^{2\pi i  m_2\Delta k_2}\nonumber\\
=n_1\delta_{\Delta k_1,0}\cdot n_2\delta_{\Delta k_2,0},
\end{gather}

\noindent where $\delta_{i,j}$ is the Kronecker symbol \cite{spiegel1997incidence,
solyom2010fundamentals}. Applying this to Eq. \eqref{eq:FT2} we get that
\begin{equation}
\sum_{\underline{m}}  d_{1}\left|B,\underline{m}\cdot\underline{\boldsymbol{a}}\right\rangle \left\langle A,\underline{m}\cdot\underline{\boldsymbol{a}}\right|=d_{1}\sum_{\boldsymbol{k}}\mathrm{e}^{-i\boldsymbol{k}\boldsymbol{\delta}_{1}}\left|B,\boldsymbol{k}\right\rangle \left\langle A,\boldsymbol{k}\right|.
\end{equation}

The Fourier transformation of the terms with different periodicity must be done
in the same manner. We show the first term of Eq. \eqref{eq:kek} as an example:

\begin{widetext}
\begin{gather}
\sum_{M_{1},M_{2}}\left|B,M_{1}\boldsymbol{C}_{1}+M_{2}\boldsymbol{C}_{2}\right\rangle \left\langle A,M_{1}\boldsymbol{C}_{1}+M_{2}\boldsymbol{C}_{2}+\boldsymbol{a}_{2}\right| \\
=\frac{1}{N}\sum_{\boldsymbol{k},\boldsymbol{k}'}\sum_{M_{1},M_{2}}\mathrm{e}^{i\boldsymbol{k}\left(M_{1}\left(2\boldsymbol{a}_{1}-\boldsymbol{a}_{2}\right)+M_{2}\left(2\boldsymbol{a}_{2}-\boldsymbol{a}_{1}\right)\right)}\left|B,\boldsymbol{k}\right\rangle \mathrm{e}^{-i\boldsymbol{k}'\left(M_{1}\left(2\boldsymbol{a}_{1}-\boldsymbol{a}_{2}\right)+M_{2}\left(2\boldsymbol{a}_{2}-\boldsymbol{a}_{1}\right)+\boldsymbol{a}_{2}\right)}\mathrm{e}^{-i\boldsymbol{k}'\boldsymbol{\delta}_{1}}\left\langle A,\boldsymbol{k}'\right| \nonumber \\
=\frac{1}{N}\sum_{\boldsymbol{k},\boldsymbol{k}'}\sum_{M_{1},M_{2}}\left|B,\boldsymbol{k}\right\rangle \left\langle A,\boldsymbol{k}'\right|\mathrm{e}^{-i\left(\boldsymbol{k}'-\boldsymbol{k}\right)\left(M_{1}\left(2\boldsymbol{a}_{1}-\boldsymbol{a}_{2}\right)+M_{2}\left(2\boldsymbol{a}_{2}-\boldsymbol{a}_{1}\right)\right)}\mathrm{e}^{-i\boldsymbol{k}'(\overbrace{\boldsymbol{a}_{2}+\boldsymbol{\delta}_{1}}^{\boldsymbol{\delta}_{2}})}.\nonumber
\end{gather}
\end{widetext}
Summing the terms that depend on $M_i$, we get:
\begin{gather}
\sum_{M_{1}=1}^{N_1}\sum_{M_{2}=1}^{N_2}\mathrm{e}^{-i\left(\boldsymbol{k}'-\boldsymbol{k}\right)\left(M_{1}\left(2\boldsymbol{a}_{1}-\boldsymbol{a}_{2}\right)+M_{2}\left(2\boldsymbol{a}_{2}-\boldsymbol{a}_{1}\right)\right)}
\end{gather}
where $N_1\cdot N_2 = N/3$ is the number of the larger unit cell respecting the
periodicity of the Kekulé pattern. We can rewrite the terms in the exponent as:
\begin{equation}
\footnotesize
M_1\left(2\boldsymbol{a}_{1}-\boldsymbol{a}_{2}\right)+M_{2}\left(2\boldsymbol{a}_{2}-\boldsymbol{a}_{1}\right)=\left(\begin{array}{cc}
\boldsymbol{a}_{1} & \boldsymbol{a}_{2}\end{array}\right)\left(\begin{array}{rr}
2 & -1\\
-1 & 2
\end{array}\right)\left(\begin{array}{c}
M_{1}\\
M_{2}
\end{array}\right).
\end{equation}
We can do the same expansion in the terms of the reciprocal lattice vectors as we
did above:
\begin{gather}
\left(\begin{array}{cc}
\Delta k_1 & \Delta k_2\end{array}\right)\left(\begin{array}{c}
\boldsymbol{b}_{1}\\
\boldsymbol{b}_{2}
\end{array}\right)\left(\begin{array}{rr}
\boldsymbol{a}_{1} & \boldsymbol{a}_{2}\end{array}\right)\left(\begin{array}{rr}
2 & -1\\
-1 & 2
\end{array}\right)\left(\begin{array}{c}
M_{1}\\
M_{2}
\end{array}\right)\nonumber\\
=2\pi\left(\begin{array}{rr}
\Delta k_1 & \Delta k_2\end{array}\right)\left(\begin{array}{rr}
2 & -1\\
-1 & 2
\end{array}\right)\left(\begin{array}{c}
M_{1}\\
M_{2}
\end{array}\right).
\end{gather}
If this exponent is a multiple of $2\pi$ then the value of the expression is 1
and 0 otherwise. Which means we can translate the question to a set of linear
equations and solve it over the integer numbers keeping in mind the fact that the
$\Delta k_i$ numbers have a finite set similarly as before. This problem has 3
different solutions: $\boldsymbol{k}-\boldsymbol{k}'=\boldsymbol{0},\pm \boldsymbol{G}$,
where {${\boldsymbol{G}=\frac{1}{3}(\boldsymbol{b}_1-\boldsymbol{b}_2)}$} is the
so-called Kekulé wave vector \cite{PhysRevB97165430,andrade2019valley}:
\begin{gather}
\sum_{M_{1}=1}^{N_1}\sum_{M_{2}=1}^{N_2}\mathrm{e}^{-i\left(\boldsymbol{k}'-\boldsymbol{k}\right)\left(M_{1}\left(2\boldsymbol{a}_{1}-\boldsymbol{a}_{2}\right)+M_{2}\left(2\boldsymbol{a}_{2}-\boldsymbol{a}_{1}\right)\right)}\\
=\frac{N}{3}\left( \delta_{\boldsymbol{k},\boldsymbol{k}'}+\delta_{\boldsymbol{k},\boldsymbol{k}'\pm\boldsymbol{G}} \right).\nonumber
\end{gather}
Now we can write down the Fourier transform:
\begin{gather}
\sum_{M_{1},M_{2}}\left|B,M_{1}\boldsymbol{C}_{1}+M_{2}\boldsymbol{C}_{2}\right\rangle \left\langle A,M_{1}\boldsymbol{C}_{1}+M_{2}\boldsymbol{C}_{2}+\boldsymbol{a}_{2}\right|\nonumber \\
=\frac{1}{3}\sum_{\boldsymbol{k}}\left|B,\boldsymbol{k}\right\rangle \left\langle A,\boldsymbol{k}\right|\mathrm{e}^{-i\boldsymbol{k}\boldsymbol{\delta}_{2}}
+\left|B,\boldsymbol{k}\right\rangle \left\langle A,\boldsymbol{k}+\boldsymbol{G}\right|\mathrm{e}^{-i\left(\boldsymbol{k}+\boldsymbol{G}\right)\boldsymbol{\delta}_{2}}\nonumber\\
+\left|B,\boldsymbol{k}\right\rangle \left\langle A,\boldsymbol{k}-\boldsymbol{G}\right|\mathrm{e}^{-i\left(\boldsymbol{k}-\boldsymbol{G}\right)\boldsymbol{\delta}_{2}}.
\end{gather}
{ 
Performing the Fourier transformation on every term in the Eq. \eqref{eq:whole2}
Hamiltonian in $\boldsymbol{k}$-space can be written as: $\hat{h}_{\uparrow/\downarrow}=\sum_{\boldsymbol{k}}\underline{\Psi}^{\dagger}_{\boldsymbol{k}}H_{\uparrow/\downarrow}(\boldsymbol{k})\underline{\Psi}_{\boldsymbol{k}}$ with

\begin{equation}
\label{eq:fullham}
H(\boldsymbol{k})_{\uparrow/\downarrow}=\left(\begin{tabular}{cc}
$\pm\boldsymbol{\Omega}$ & $\boldsymbol{\Gamma}$ \\
$\boldsymbol{\Gamma}^\dagger$ & $\mp\boldsymbol{\Omega}^\dagger$ \\
\end{tabular} \right),
\end{equation}
}
where
\begin{widetext}
\begin{equation}
\underline{\Psi}^{\dagger}_{\boldsymbol{k}}=\left(\begin{array}{cccccc}
\left|B\boldsymbol{k}\right\rangle  & \left|B,\boldsymbol{k}+\boldsymbol{G}\right\rangle  & \left|B,\boldsymbol{k}-\boldsymbol{G}\right\rangle  & \left|A,\boldsymbol{k}\right\rangle  & \left|A,\boldsymbol{k}+\boldsymbol{G}\right\rangle  & \left|A,\boldsymbol{k}-\boldsymbol{G}\right\rangle \end{array}\right),
\end{equation}
\begin{equation}
\boldsymbol{\Gamma}=\left(\begin{tabular}{ccc}

$\left(t+\frac{2\Delta}{3}\right)s(\boldsymbol{k},\boldsymbol{0})/2$ & $\frac{\Delta}{3}s(\boldsymbol{k},\boldsymbol{G})$ & $\frac{\Delta}{3}s(\boldsymbol{k},-\boldsymbol{G})$ \\
$\frac{\Delta}{3}s(\boldsymbol{k}+\boldsymbol{G},-\boldsymbol{G})$ & $\left(t+\frac{2\Delta}{3}\right)s(\boldsymbol{k}+\boldsymbol{G},\boldsymbol{0})/2$ & $\frac{\Delta}{3}s(\boldsymbol{k}+\boldsymbol{G},\boldsymbol{G})$ \\
$\frac{\Delta}{3}s(\boldsymbol{k}-\boldsymbol{G},\boldsymbol{G})$ & $\frac{\Delta}{3}s(\boldsymbol{k}-\boldsymbol{G},-\boldsymbol{G})$ & $\left(t+\frac{2\Delta}{3}\right)s(\boldsymbol{k}-\boldsymbol{G},\boldsymbol{0})/2$ \\
\end{tabular} \right),
\end{equation}

\begin{equation}
\boldsymbol{\Omega}=\frac{im}{3\sqrt{3}}\left(\begin{tabular}{ccc}
$s'\left( \boldsymbol{k},\boldsymbol{0} \right)-s'^*\left( \boldsymbol{k},\boldsymbol{0} \right)$ &  $s'\left( \boldsymbol{k},\boldsymbol{G} \right)-s'^*\left( \boldsymbol{k}+\boldsymbol{G},-\boldsymbol{G} \right)$ & $s'\left( \boldsymbol{k},-\boldsymbol{G} \right)-s'^*\left( \boldsymbol{k}-\boldsymbol{G},\boldsymbol{G} \right)$ \\
$s'\left( \boldsymbol{k}+\boldsymbol{G},-\boldsymbol{G} \right)-s'^*\left( \boldsymbol{k},\boldsymbol{G} \right)$&$s'\left( \boldsymbol{k}+\boldsymbol{G},\boldsymbol{0} \right)-s'^*\left( \boldsymbol{k}+\boldsymbol{G},\boldsymbol{0} \right)$& $s'\left( \boldsymbol{k}+\boldsymbol{G},\boldsymbol{G} \right)-s'^*\left( \boldsymbol{k}-\boldsymbol{G},-\boldsymbol{G} \right)$ \\
$s'\left( \boldsymbol{k}-\boldsymbol{G},\boldsymbol{G} \right)-s'^*\left( \boldsymbol{k},-\boldsymbol{G} \right)$&$s'\left( \boldsymbol{k}-\boldsymbol{G},-\boldsymbol{G} \right)-s'^*\left( \boldsymbol{k}+\boldsymbol{G},\boldsymbol{G} \right)$&$s'\left( \boldsymbol{k}-\boldsymbol{G},\boldsymbol{0} \right)-s'^*\left( \boldsymbol{k}-\boldsymbol{G},\boldsymbol{0} \right)$
\end{tabular} \right).
\end{equation}
\end{widetext}

Here we introduced the following functions:
\begin{gather}
s\left(\boldsymbol{k},\boldsymbol{p}\right)=d_{1}\mathrm{e}^{-i\boldsymbol{k}\left(1+\boldsymbol{\varepsilon}\right)\tilde{\boldsymbol{\delta}}_{1}}\left(\mathrm{e}^{-i\boldsymbol{p}\tilde{\boldsymbol{\delta}}_{3}}+\mathrm{e}^{-i\boldsymbol{p}\tilde{\boldsymbol{\delta}}_{2}}\right)\\
+d_{2}\mathrm{e}^{-i\boldsymbol{k}\left(1+\boldsymbol{\varepsilon}\right)\tilde{\boldsymbol{\delta}}_{2}}\left(\mathrm{e}^{-i\boldsymbol{p}\tilde{\boldsymbol{\delta}}_{1}}+\mathrm{e}^{-i\boldsymbol{p}\tilde{\boldsymbol{\delta}}_{2}}\right)\nonumber\\
+d_{3}\mathrm{e}^{-i\boldsymbol{k}\left(1+\boldsymbol{\varepsilon}\right)\tilde{\boldsymbol{\delta}}_{3}}\left(\mathrm{e}^{-i\boldsymbol{p}\tilde{\boldsymbol{\delta}}_{3}}+\mathrm{e}^{-i\boldsymbol{p}\tilde{\boldsymbol{\delta}}_{1}}\right),\nonumber
\end{gather}

\begin{gather}
s'\left(\boldsymbol{k},\boldsymbol{p}\right)=d_{4}\mathrm{e}^{-i\boldsymbol{k}\left(1+\boldsymbol{{\varepsilon}}\right)\tilde{\boldsymbol{a}}_{1}}\mathrm{e}^{-i\boldsymbol{p}\tilde{\boldsymbol{a}}_{1}}\\
+d_{5}\mathrm{e}^{i\boldsymbol{k}\left(1+\boldsymbol{{\varepsilon}}\right)\tilde{\boldsymbol{a}}_{2}}
+d_{6}\mathrm{e}^{-i\boldsymbol{k}\left(1+\boldsymbol{{\varepsilon}}\right)\left(\tilde{\boldsymbol{a}}_{2}-\tilde{\boldsymbol{a}}_{1}\right)}\mathrm{e}^{-i\boldsymbol{p}\tilde{\boldsymbol{a}}_{2}}.\nonumber
\end{gather}

In order to obtain the desired low-energy approximation of the \eqref{eq:fullham} Hamiltonian we neglect the terms that correspond to the high energy bands. The remaining 4 states can be reordered in the vector of states following the convention of Andrade \emph{et.\:al} \cite{andrade2019valley} as:

\begin{equation}
\tilde{\underline{\Psi}}_{\boldsymbol{k}}=
\left(\begin{array}{r}
-\left|A,\boldsymbol{k}-\boldsymbol{G}\right\rangle\\[0.06em]
\left|B,\boldsymbol{k}-\boldsymbol{G}\right\rangle \\ [0.06em]
\left|B,\boldsymbol{k}+\boldsymbol{G}\right\rangle \\[0.06em]
\left|A,\boldsymbol{k}+\boldsymbol{G}\right\rangle
\end{array}
\right).
\end{equation}

Strain alters the hopping energies as it was introduced in Eqs. \eqref{eq:strain_hopping} and \eqref{eq:strain_hopping2}. We expand the corresponding factors up to the first order in strain:

\begin{gather}
{d_{i}}\approx 1-\beta\left(\frac{|\left(\boldsymbol{\varepsilon}+\boldsymbol{I}\right)\tilde{\boldsymbol{{\delta}}}_{i}|}{a_\mathrm{cc}}-1\right)\\
=1-\beta\left(\frac{\sqrt{\left(\boldsymbol{\varepsilon}\tilde{\boldsymbol{{\delta}}}_{i}+\tilde{\boldsymbol{{\delta}}}_{i}\right)^{\mathrm{T}}\left(\boldsymbol{\varepsilon}\tilde{\boldsymbol{{\delta}}}_{i}+\tilde{\boldsymbol{{\delta}}}_{i}\right)}}{a_\mathrm{cc}}-1\right)\nonumber\\
\approx1-\frac{\beta}{a_\mathrm{cc}^{2}}\tilde{\boldsymbol{\delta}}^{\mathrm{T}}_{i}\boldsymbol{\varepsilon}\tilde{\boldsymbol{\delta}}_{i}\nonumber.
\end{gather}

Next we proceed to expand \eqref{eq:fullham} up to first order in $\boldsymbol{k}$. To this end we can make an expansion of $s$ and $s'$ functions around $\boldsymbol{G}$. However, as other works already have shown, it is necessary to expand around the true Dirac points, which are defined as the zeros of the deformed lattice energy dispersion \cite{oliva2013understanding,andrade2019valley,botello2018toward,naumis2017electronic}. Generally these are located neither at the high-symmetry points of the strained lattice nor at the tips of the original Dirac cones. These new $\boldsymbol{k}$-points are given by $\boldsymbol{K}=\pm(\boldsymbol{G}+\boldsymbol{A})$. The components of the pseudovector-potential $\boldsymbol{A}$ can be expressed with the matrix element of $\boldsymbol{\varepsilon}$ such as \cite{vozmediano2010gauge}:

\begin{gather}
A_x = \frac{\beta}{2a_{\mathrm{cc}}}\left( \varepsilon_{xx}-\varepsilon_{yy} \right), \\ \nonumber
A_y = -\frac{\beta}{2a_{\mathrm{cc}}}\left( 2\varepsilon_{xy} \right).
\end{gather}

Applying these approximations to our Hamiltonian \eqref{eq:fullham} after some straightforward but slightly tedious algebra we get the following low-energy Hamiltonian:

\begin{widetext}
\begin{gather}
H_\mathrm{low-energy}\left(\boldsymbol{k}\right)=-\frac{3}{2}a_\mathrm{cc}\left(t+\frac{2\Delta}{3}\right)\left[\boldsymbol{\sigma}\boldsymbol{A}\otimes\tau_{z}+ \boldsymbol{\sigma}\left(\boldsymbol{1}+\left(1-\beta\right)\boldsymbol{\varepsilon}\right)\boldsymbol{k}\otimes\tau_{0}\right]\\
-\frac{\Delta}{2}\left[\left(\beta\mathrm{Tr}\left(\boldsymbol{\varepsilon}\right)-2\right)\sigma_{z}\otimes\tau_{x}-
a_\mathrm{cc}^2(\boldsymbol{A}\times\boldsymbol{k})_{z}\sigma_{0}\otimes\tau_{y}+a_\mathrm{cc}^2\boldsymbol{A}\boldsymbol{k}\sigma_{0}\otimes\tau_{x}\right] \nonumber\\
+\frac{m}{2}\left[ \boldsymbol{A}\boldsymbol{k}\sigma_z\otimes\tau_z-\left(\frac{\beta}{2}\mathrm{Tr}\left(\boldsymbol{\varepsilon} \right)-1+3\boldsymbol{A}\boldsymbol{k} \right)\sigma_0\otimes\tau_z \right]+m\left[ \boldsymbol{A}+\boldsymbol{k}+\left(1-\frac{\beta}{2} \right)\boldsymbol{\varepsilon}\boldsymbol{k}-\frac{\beta}{4}\mathrm{Sp}\left(\boldsymbol{\varepsilon} \right)\boldsymbol{k} \right]\sigma_x\otimes\boldsymbol{\tau'},\nonumber
\end{gather}
\end{widetext}

where $\boldsymbol{\tau'}=\left(\tau_x,-\tau_y \right)$.
This formula can be significantly simplified if only small perturbations in $m$ and $\Delta$ are considered keeping only the linear terms in $m$, $\Delta$ and $\boldsymbol{k}$ and neglecting all the multilinear contributions. With the inclusion of the spin degree of freedom our effective Hamiltonian reads:

\begin{gather}
H_\mathrm{eff} =-v\left[\boldsymbol{\sigma}\boldsymbol{A}\otimes\tau_{z}+\boldsymbol{\sigma}\boldsymbol{k}\otimes\tau_{0}\right]\otimes s_0 \\ \nonumber
+\Delta\cdot\sigma_{z}\otimes\tau_{x}\otimes s_0-m\cdot \sigma_{z}\otimes\tau_{0}\otimes s_z,
\end{gather}

which is equivalent to our simplified effective Hamiltonian in Section \ref{model} at Eq.\:\eqref{eq:1}.

{

\section{}
\begin{figure}
\includegraphics[width=1\columnwidth]{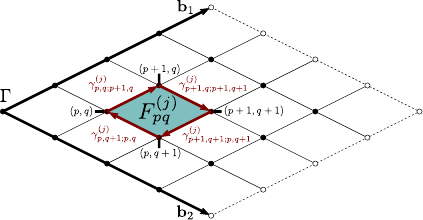}
\caption{ {
Discretization of the Brillouin zone spanned by reciprocal lattice vectors
$\boldsymbol{b}_1$ and $\boldsymbol{b}_2$. Each vertex denoted by a black dot and labeled
by indecies $(n,m)$ is associated to a plaquette. The Berry flux, $F^{(j)}_{nm}$,
is calculated for each band on each plaquette. White dots depict vertices whose
plaquettes are located in the adjacent Brillouin zone.
(color online)}}
\label{fig:5abra}
\end{figure}

\begin{figure}
\includegraphics[width=1\columnwidth]{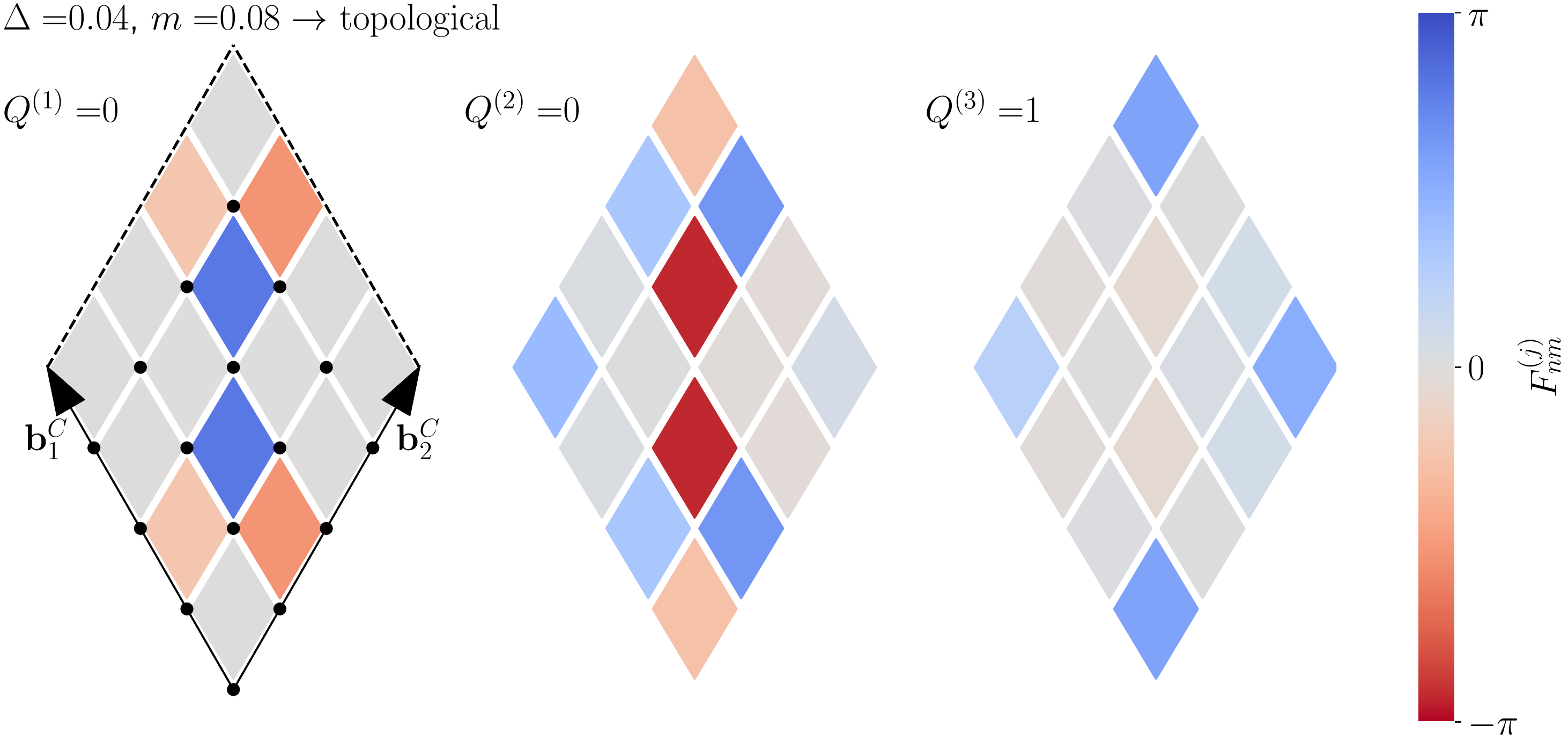}
\newline
\newline
\includegraphics[width=1\columnwidth]{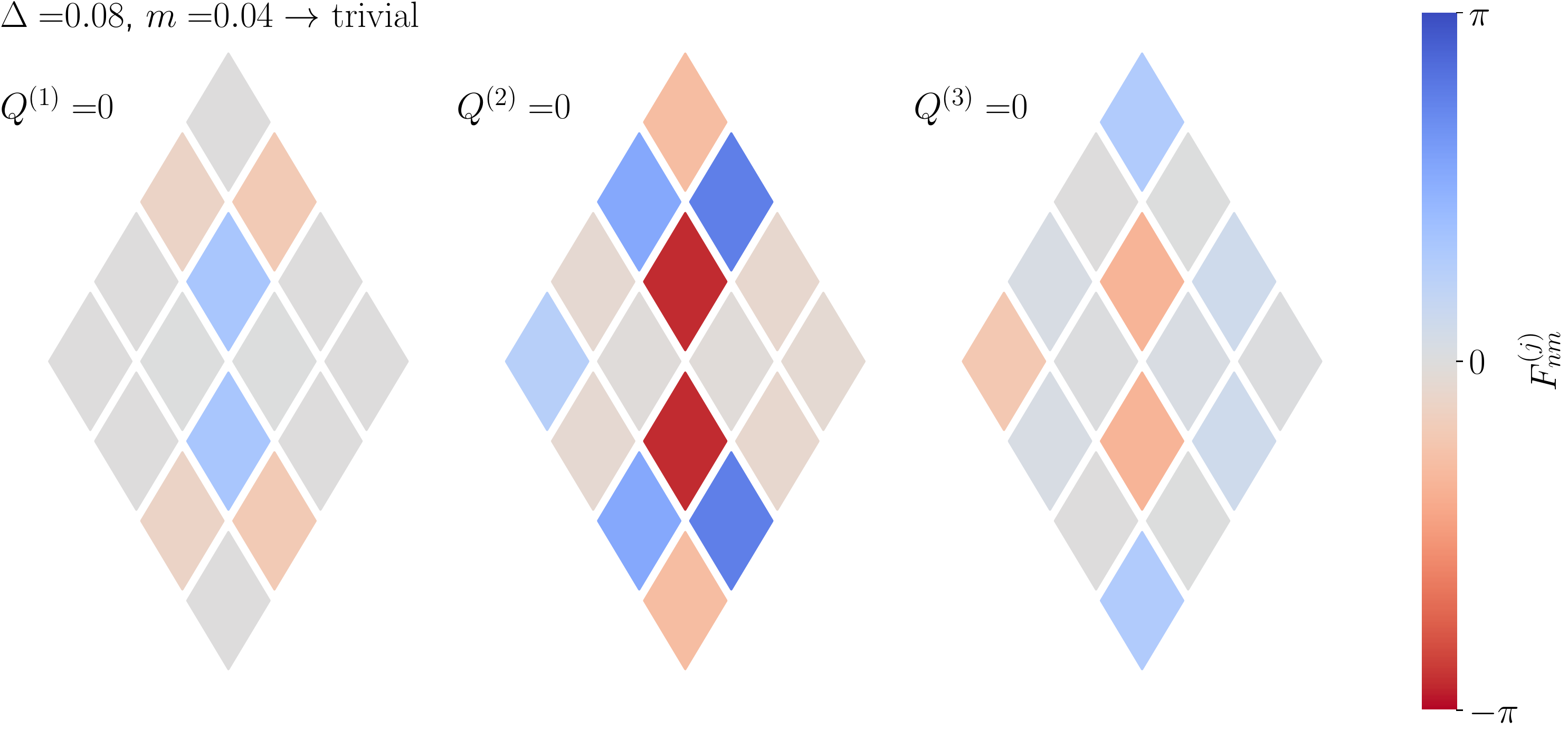}
\caption{ {Calculated Chern numbers and the distribution of the
Berry flux over the Brillouin zone in the absence of strain for the three occupied bands for topological and trivial phases.  \\
(color online)}}
\label{fig:6abra}
\end{figure}

In this section we outline the calculation of the topological invariant of the
investigated system. As the considered spin-orbit coupling preserves the $z$
component of the spin the $\mathbb{Z}_2$ invariant, characterizing time reversal
invariant models, can be obtained from the parity of the total Chern number of
the occupied band calculated for one of the spin species
\cite{Fukui2005,asboth_short_2016}.

Consider a discretization of the Brillouin zone depicted in Fig.~\ref{fig:5abra}.
Assuming that the Brillouin is spanned by lattice vectors $\boldsymbol{b}_1$ and
$\boldsymbol{b}_2$ we introduce sampling points labeled by $(p,q)$ corresponding
to wavenumbers $\boldsymbol{k}=\frac{p\boldsymbol{b}_1}{N}+\frac{q\boldsymbol{b}_2}{N}$ with
$N$ being a positive integer and $p,q\in[0,\dots,N-1]$. In Fig.~\ref{fig:5abra}
these points are depicted by black dots.
For each vertex $(p,q)$ we denote the $j-$th eigenstate of the system by $|j;(p,q)\rangle$
For each vertex $(p,q)$ we associate a plaquette.
Assuming that the spectrum of the system is non degenerate over the entire
Brillouin zone the Chern number $Q^{(j)}$ for the $j-$th occupied band is the sum
of Berry fluxes $F^{(j)}_{pq}$ for each plaquette
\begin{equation}
Q^{(j)}=\frac{1}{2\pi}\sum_{pq}F^{(j)}_{pq}.
\end{equation}
The Berry fluxes $F^{(j)}_{pq}$ are in turn determined by the four phase angles \begin{equation}
\gamma^{(j)}_{p,q;p',q'} = -\mathrm{arg}\left( \big\langle  j;(p,q) \big | j;(p',q') \big\rangle\right),
\end{equation}
defined on the edge bonds, denoted by red arrows in Fig.~\ref{fig:5abra},  corresponding to the given plaquette:

\begin{align}
F_{pq}^{(j)}=-\text{arg}\left  ( \exp \left ( -\text{i} \left [ \gamma_{p,q;p+1,q}^{(j)}+\gamma_{p+1,q;p+1,q+1}^{(j)}\right . \right . \right . \\
\nonumber \left . \left . \left .  + \gamma_{p+1,q+1;p,q+1}^{(j)} + \gamma_{p,q+1;p,q}^{(j)} \right ] \right ) \right  )
\end{align}

Assuming that the spectrum of the system is non degenerate the Chern number for
a given band is obtained by the sum of the Berry fluxes for each plaquette in
the Brillouin zone. The total Chern number $Q$ is the sum of the Chern numbers
$Q^{(j)}$ of the occupied bands while the parity of $Q$ gives the topological
invariant for the system investigated in the present manuscript. Even/odd Chern
numbers correspond to a trivial/tropological phase.

As an illustration of the procedure we apply it to $H(\boldsymbol{k})$ defined
in \eqref{eq:fullham}. First we note that the real periodicity of our model is
dictated by the periodicity of the modulation characterized by the real space
vectors $\boldsymbol{C}_1$ and $\boldsymbol{C}_2$ we shall denote the corresponding
reciprocal lattice vectors by $\boldsymbol{b}^C_1$ and $\boldsymbol{b}^C_2$. In
Fig.~\ref{fig:6abra} we depict the calculated Berry fluxes and Chern numbers
of the three lower bands, we considered as occupied, for a topological and for
a trivial case. As one can observe already a modest $N=4$ discretization yields
the right topological character for the investigated system.

}

\bibliography{cikkek}

\end{document}